\documentclass[conference,10pt]{IEEEtran}
\IEEEoverridecommandlockouts

\usepackage{epsfig,rotating,setspace,latexsym,amsmath,epsf,amssymb,amsfonts,bm,theorem,cite,algorithm,graphicx,epsf,authblk,epstopdf,color,algpseudocode,bbm}

\allowdisplaybreaks

\begin{document}

\title{Securing Downlink Non-Orthogonal Multiple Access Systems by Trusted Relays\thanks{This research was supported in part by the U.S. National Science
Foundation under Grants ECCS 1549881 and ECCS 1647198.}}

\author[1]{Ahmed Arafa}
\author[2,1]{Wonjae Shin}
\author[3,1]{Mojtaba Vaezi}
\author[1]{H. Vincent Poor}
\affil[1]{\normalsize Electrical Engineering Department, Princeton University}
\affil[2]{\normalsize Department of Electronics Engineering, Pusan National University}
\affil[3]{\normalsize Electrical and Computer Engineering Department, Villanova University}

\maketitle

%================================
\begin{abstract}
A downlink single-input single-output non-orthogonal multiple access system is considered in which a base station (BS) is communicating with two legitimate users in the presence of an external eavesdropper. A group of trusted cooperative half-duplex relay nodes, powered by the BS, is employed to assist the BS's transmission. The goal is to design relaying schemes such that the legitimate users' {\it secrecy rate region} is maximized subject to a total power constraint on the BS and the relays' transmissions. Three relaying schemes are investigated: {\it cooperative jamming}, {\it decode-and-forward}, and {\it amplify-and-forward}. Depending on the scheme, {\it secure beamforming} signals are carefully designed for the relay nodes that either diminish the eavesdropper's rate without affecting that of the legitimate users, or increase the legitimate users' rates without increasing that of the eavesdropper. The results show that there is no relaying scheme that fits all conditions; the best relaying scheme depends on the system parameters, namely, the relays' and eavesdropper's distances from the BS, and the number of relays. They also show that the relatively simple cooperative jamming scheme outperforms other schemes when the relays are far from the BS and/or close to the eavesdropper.
\end{abstract}

%================================
\section{Introduction}

Non-orthogonal multiple access (NOMA) systems are leading candidates for future mobile communications, as they provide an efficient means of utilizing resources serving multiple users simultaneously compared to orthogonal multiple access schemes \cite{poor-noma-intro}. With the inherent broadcast nature of wireless communications, securing transmitted data is critical in the presence of potential eavesdroppers in the medium. Physical layer security is a powerful, and provably unbreakable, technique to deliver secure data as opposed to conventional security techniques implemented in higher layers of the communication protocol stack; see, e.g., \cite{poor-wireless-pls} and the references therein. In this work, we focus on enhancing physical layer security of single-input single-output (SISO) NOMA systems by using cooperative relays.

There have been a number of recent works that study physical layer security for NOMA systems \cite{ding-noma-sum-sec-rate, hanzo-noma-sec-large-scale-downlink, elkashlan-noma-sec-large-scale-uplink, ding-noma-unicast-multicast, qin-noma-sum-sec-eav-mimo, qin-noma-sum-sec-edge-miso, qin-noma-sum-sec-edge-mimo, alouini-noma-sec-antenna-select, lau-noma-outage}. Secrecy sum rate maximization of SISO NOMA systems is studied in \cite{ding-noma-sum-sec-rate}. Using tools from stochastic geometry, references \cite{hanzo-noma-sec-large-scale-downlink} and \cite{elkashlan-noma-sec-large-scale-uplink} study security measures for large scale NOMA systems in the downlink and uplink, respectively. NOMA assisted multicast-unicast streaming is studied in \cite{ding-noma-unicast-multicast}, where secure rates for unicast transmission using NOMA is shown to outperform conventional orthogonal schemes. Reference \cite{qin-noma-sum-sec-eav-mimo} considers a multiple-input multiple-output (MIMO) two-user NOMA setting with an external eavesdropper and designs beamforming signals that maximize the secrecy sum rate. This approach is also considered in \cite{qin-noma-sum-sec-edge-miso} and \cite{qin-noma-sum-sec-edge-mimo} for multiple-input single output and MIMO scenarios, respectively, with the assumption that one user is entrusted and the other is the potential eavesdropper. The impact of transmit antenna selection strategies on the secrecy outage probability is investigated in \cite{alouini-noma-sec-antenna-select}. Transmit power minimization and minimum secrecy rate maximization subject to a secrecy outage constraint is considered in \cite{lau-noma-outage}.

\begin{figure}[t]
\center
\includegraphics[scale=.95]{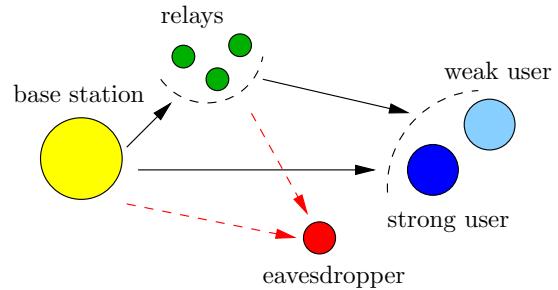}
\caption{Downlink NOMA system model, with cooperative relays, and an external eavesdropper.}
\label{fig_sys_model}
\vspace{-.25in}
\end{figure}

Our work is most closely related to \cite{petropulu-coop-relay-security}, where relays are used to enhance the secrecy rate of a single-receiver wiretap channel. In this paper, we extend the ideas in \cite{petropulu-coop-relay-security} to work in the context of a two-user downlink SISO NOMA system with an external eavesdropper. We employ multiple {\it trusted cooperative} half-duplex relays to enhance the achievable secrecy rate region through various relaying schemes: {\it cooperative jamming}, {\it decode-and-forward}, and {\it amplify-and-forward}. For each scheme, we design {\it secure beamforming} signals at the relays that benefit the users and/or hurt the eavesdropper. Under a total system power constraint, which is divided between the base station (BS) and the relays, an achievable secrecy rate region for each relaying scheme is derived and analyzed. In general, the results show that the best relaying scheme highly depends on the system parameters, in particular the distances between nodes, and that the relatively simple cooperative jamming scheme performs better than the other schemes when the relays are close to the eavesdropper.

%================================
\section{System Model}

We consider a two-user downlink SISO NOMA system in which the BS uses superposition coding to send messages to the two users simultaneously. User channels are fixed during the communication session and are known at the BS. The user with a relatively worse channel condition (weak user) decodes its message by treating the other user's interfering signal as noise, while the user with a relatively better channel condition (strong user) first decodes the weak user's message, by treating its own signal as noise, and then uses successive interference cancellation to decode its own message\footnote{The two-user setting in this work is adopted in NOMA systems in which users are divided into multiple clusters with two users each, in order to reduce error propagation in successive interference cancellation decoding \cite{poor-noma-intro}.}.

We denote the channel between the BS and the strong user (resp. weak user) by $h_1$ (resp. $h_2$), with\footnote{All channel gains in this paper are complex-valued, and are drawn independently from some continuous distribution.} $|h_1|^2>|h_2|^2$. The received signals at the strong and weak users are given by
\begin{align}
y_1&=h_1x+n_1 \label{eq_rec_sig_1} \\
y_2&=h_2x+n_2 \label{eq_rec_sig_2}
\end{align}
where the noise terms $n_1$ and $n_2$ are independent and identically distributed (i.i.d.) circularly-symmetric complex Gaussian random variables with zero means and unit variances, $\mathcal{CN}\left(0,1\right)$, and the transmitted signal $x$ is given by
\begin{align} \label{eq_trans_sgnl}
x=\sqrt{\alpha P}s_1+\sqrt{\bar{\alpha}P}s_2
\end{align}
where $s_1$ and $s_2$ are i.i.d. $\sim\mathcal{CN}\left(0,1\right)$ information carrying signals for the strong and the weak user, respectively, $P$ is the BS's transmit power budget, $\alpha\in[0,1]$ is the fraction of power allocated to the strong user, and $\bar{\alpha}\triangleq1-\alpha$. Using superposition coding and successive interference cancellation decoding, one achieves the following rates of this (degraded) Gaussian broadcast channel \cite{cover}:
\begin{align}
r_1&=\log\left(1+|h_1|^2\alpha P\right) \label{eq_rate_1}\\
r_2&=\log\left(1+\frac{|h_2|^2\bar{\alpha}P}{1+|h_2|^2\alpha P}\right).\label{eq_rate_2}
\end{align}

An external eavesdropper is monitoring the communication between the BS and the users. We denote the channel between the BS and the eavesdropper by $h_e$. Thus, the received signal at the eavesdropper is given by
\begin{align}
y_e=h_ex+n_e \label{eq_rec_sig_e}
\end{align}
where the noise term $n_e\sim\mathcal{CN}\left(0,1\right)$. For a given $0\leq\alpha\leq1$, the secrecy capacities of the two users of this multi receiver wiretap channel is given by \cite[Theorem 5]{ersen-mr-wt}
\begin{align}
r_{s,1}&\!=\!\left[\log\left(1+|h_1|^2\alpha P\right)-\log\left(1+|h_e|^2\alpha P\right)\right]^+ \label{eq_sec_rate_1} \\
r_{s,2}&\!=\!\left[\log\left(\!1\!+\!\frac{|h_2|^2\bar{\alpha}P}{1\!+\!|h_2|^2\alpha P}\!\right)\!\!-\!\log\!\left(\!1\!+\!\frac{|h_e|^2\bar{\alpha}P}{1\!+\!|h_e|^2\alpha P}\!\right)\right]^+ \label{eq_sec_rate_2}
\end{align}
where the subscript $s$, here and throughout the paper, is to denote secrecy rates, and $[x]^+\triangleq\max\{x,0\}$.

It is clear from (\ref{eq_sec_rate_1}) and (\ref{eq_sec_rate_2}) that sending secure data depends on the eavesdropper's channel condition with respect to that of the legitimate users. Therefore, we propose using trusted cooperative half-duplex {\it relay} nodes, see Fig.~\ref{fig_sys_model}, to assist the BS via three possible schemes: {\it cooperative jamming}, {\it decode-and-forward}, and {\it amplify-and-forward}. In all of these schemes, the BS uses only a portion of its available power $\bar{P}\leq P$ for its own transmission, and shares the remaining portion $P-\bar{P}$ with the relays for their transmission. We discuss these schemes in detail in the following sections, after introducing the relays' channels notation that we use.

Let there be $K$ relays, and denote the channel gains from the BS to the relays by the vector\footnote{All vectors in this paper are column vectors.} ${\bm h}_r\triangleq[h_{r,1},\dots,h_{r,K}]$. Let ${\bm g}_1$, ${\bm g}_2$, and ${\bm g}_e$ denote the $K$-length channel gain vectors from the relays to the first user, the second user, and the eavesdropper, respectively. The channels from the relays to the users and the eavesdropper are known at the relays.

%================================
\section{Cooperative Jamming} \label{sec_coop_jam}

In this section, we discuss the cooperative jamming scheme. {\it Simultaneously} with the BS's transmission, the relays transmit a cooperative jamming signal ${\bm J}z$, where ${\bm J}\in\mathbb{C}^K$ is a beamforming vector and $z\sim\mathcal{CN}(0,1)$. The received signals at the users and the eavesdropper are in this case given by
\begin{align}
y_j&=h_jx+{\bm g}_j^\dagger{\bm J}z+n_j,\quad j=1,2,e
\end{align}
%\begin{align}
%y_1&=h_1x+{\bm g}_1^\dagger{\bm J}z+n_1 \\
%y_2&=h_2x+{\bm g}_2^\dagger{\bm J}z+n_2 \\
%y_e&=h_ex+{\bm g}_e^\dagger{\bm J}z+n_e
%\end{align}
where the superscript $\dagger$ denotes the conjugate transpose operation. The signal power is now set to $\mathbb{E}\left[|x|^2\right]=\bar{P}\leq P$, where $\mathbb{E}[\cdot]$ denotes expectation, and the remaining portion $P-\bar{P}$ is used to power the relays, i.e., $\mathbb{E}\left[\left({\bm J}z\right)^\dagger\left({\bm J}z\right)\right]={\bm J}^\dagger{\bm J}=P-\bar{P}$.

The design of the jamming signal is such that it has minimal effect on the legitimate users as follows:
\begin{align} \label{eq_cj_null}
\left[{\bm g}_1 ~~~ {\bm g}_2\right]^\dagger{\bm J}\triangleq{\bm G}^\dagger{\bm J}=\left[0~~~0\right].
\end{align}
That is, we choose ${\bm J}$ in the null space of the matrix ${\bm G}^\dagger$. If there exist $K\geq3$ relays, then ${\bm G}^\dagger$ will always have a nonempty null space and (\ref{eq_cj_null}) will have a nontrivial solution. We denote such a null space jamming signal by ${\bm J}_o$. Since the channel state vectors ${\bm g}_1$, ${\bm g}_2$, and ${\bm g}_e$ are drawn from a continuous distribution, they are therefore linearly independent with probability $1$ (w.p. $1$). Thus, we have
\begin{align}
\left|{\bm g}_e^\dagger{\bm J}_o\right|>0,\quad\text{w.p. $1$}.
\end{align}
Therefore, the achievable secrecy rates are now given by (\ref{eq_sec_rate_1}) and (\ref{eq_sec_rate_2}) after replacing $h_e$ with $\tilde{h}_e$ which is given by
\begin{align} \label{eq_he_tilde}
\tilde{h}_e\triangleq h_e\Big/\left(1+\left|{\bm g}_e^\dagger{\bm J}_o\right|^2\right).
\end{align}

We now find the optimal ${\bm J}_o$ that maximally degrades the eavesdropper's channel subject to not affecting the legitimate users' channels. Observe that both legitimate users' rates are decreasing in $\tilde{h}_e$, i.e., increasing in $\left|{\bm g}_e^\dagger{\bm J}_o\right|^2$. Therefore, to maximize their secrecy rates, we formulate the following optimization problem for a given transmit power $\bar{P}$:
\begin{align}
\max_{{\bm J}_o} \quad &\left|{\bm g}_e^\dagger{\bm J}_o\right|^2 \nonumber \\
\mbox{s.t.} \quad &{\bm G}^\dagger{\bm J}_o=\begin{bmatrix}0&0\end{bmatrix},~{\bm J}_o^\dagger{\bm J}_o=P-\bar{P}.
\end{align}
The above problem has a unique solution \cite{friedlander-null-steering} (see also \cite{petropulu-coop-relay-security}), which we derive next for completeness. We first rewrite the first constraint slightly differently as follows:
\begin{align}
{\bm J}_o=\mathcal{P}^\perp\!\left({\bm G}\right){\bm u}_J
\end{align}
for some vector ${\bm u}_J\in\mathbb{C}^K$ to be designed, and $\mathcal{P}^\perp\!\left({\bm G}\right)$ is the orthogonal projection matrix onto the null space of ${\bm G}^\dagger$:
\begin{align} \label{eq_p_perp_original}
\mathcal{P}^\perp\!\left({\bm G}\right)\triangleq{\bm I}_K-{\bm G}\left({\bm G}^\dagger{\bm G}\right)^{-1}{\bm G}^\dagger
\end{align}
where ${\bm I}_K$ is the $K$-dimensional identity matrix. It is now direct to see that the vector ${\bm u}_J$ should be chosen along the same direction of $\mathcal{P}^\perp\!\left({\bm G}\right){\bm g}_e$ in order to maximize $\left|{\bm g}_e^\dagger{\bm J}_o\right|^2$. Finally, to satisfy the power constraint, the optimal beamforming vector, $\hat{{\bm J}}_o$, is given by
\begin{align}
\hat{{\bm J}}_o=\frac{\mathcal{P}^\perp\!\left({\bm G}\right){\bm g}_e}{\left\|\mathcal{P}^\perp\!\left({\bm G}\right){\bm g}_e\right\|}\sqrt{P-\bar{P}}
\end{align}
which, upon substituting in (\ref{eq_he_tilde}), achieves the following secrecy rates for a given $\bar{P}$ and $\alpha$:
\begin{align}
&r_{s,1}^J\!=\!\left[\log\!\left(1+|h_1|^2\alpha \bar{P}\right)\!\right. \nonumber \\
&\left.\hspace{.625in}-\!\log\!\left(1+\frac{|h_e|^2\alpha \bar{P}}{1+{\bm g}_e^\dagger\mathcal{P}^\perp\!\left({\bm G}\right){\bm g}_e\left(P-\bar{P}\right)}\right)\!\right]^+ \\
&r_{s,2}^J\!=\!\left[\log\left(1+\frac{|h_2|^2\bar{\alpha}\bar{P}}{1+|h_2|^2\alpha \bar{P}}\right)\right. \nonumber \\
&\left.\hspace{-.0625in}-\log\left(1+\frac{|h_e|^2\bar{\alpha}\bar{P}}{1+|h_e|^2\alpha \bar{P}+{\bm g}_e^\dagger\mathcal{P}^\perp\!\left({\bm G}\right){\bm g}_e\left(P-\bar{P}\right)}\right)\right]^+ 
\end{align}
with the superscript $J$ denoting cooperative jamming.

In Section~\ref{sec_num}, we discuss the evaluation of the optimal transmit power $\bar{P}$ and the power fraction $\alpha$ that maximize the secrecy rate region of this cooperative jamming scheme, along with those that maximize the secrecy rate regions of the other relaying schemes that we consider in this paper.

%================================
\section{Decode-and-Forward} \label{sec_dec_fwd}

In this section, we discuss the decode-and-forward scheme. Different from the cooperative jamming scheme, communication takes place over two phases. In the first phase, the BS broadcasts the messages to both the relays and the legitimate users. In the second phase, the relays forward the messages that they decoded to the legitimate users. The eavesdropper overhears the communication during both phases.

The received signals during the first phase at the legitimate users and the eavesdropper are given by (\ref{eq_rec_sig_1})--(\ref{eq_rec_sig_2}) and (\ref{eq_rec_sig_e}), respectively, with a total transmit power $\bar{P}\leq P$. The received signals at the relays during the first phase is given by
\begin{align} \label{eq_rec_sig_r}
{\bm y}_r={\bm h}_rx+{\bm n}_r
\end{align}
where the noise term vector ${\bm n}_r\sim\mathcal{CN}\left(0,{\bm I}_K\right)$. During the first phase, each relay decodes the users' messages by either decoding the weak user's message first by treating the strong user's signal as noise, and then decoding the strong user's message via successive interference cancellation, or it can do the whole process in the reverse order by decoding the strong user's message first followed by that of the weak user. Choosing the decoding order depends on in which part of the rate region the system is operating. For simplicity, we let all the relays choose the same decoding order, i.e., they all take one collective decision on that matter. Let the superscript $(i)$, $i=1,2$, denote the $i$th decoding order, with $i=1$ denoting decoding the strong user's message first followed by that of the weak user, and $i=2$ denoting the reverse order. Thus, the achieved rate pairs at the $k$th relay after the first phase are either given as follows for $i=1$: \newline $\left(R_{k,1}^{(1)},R_{k,2}^{(1)}\right)\!=\!\left(\!\log\left(1+\frac{|h_{r,k}|^2\alpha\bar{P}}{1+|h_{r,k}|^2\bar{\alpha}\bar{P}}\!\right),\log\left(1+|h_{r,k}|^2\bar{\alpha}\bar{P}\right)\right)$, or are given as follows for $i=2$: \newline $\left(R_{k,1}^{(2)},R_{k,2}^{(2)}\right)\!=\!\left(\!\log\left(1+|h_{r,k}|^2\alpha\bar{P}\right),\log\left(1+\frac{|h_{r,k}|^2\bar{\alpha}\bar{P}}{1+|h_{r,k}|^2\alpha\bar{P}}\right)\!\right)$.

In the second phase, the relays form the transmitted signal $x_r$, which is exactly as in (\ref{eq_trans_sgnl}) but after replacing $P$ with $P-\bar{P}$. We assume that the relays use the same power fraction $\alpha$ in the second phase. While in general each relay can use a different power fraction, we use the same fraction for simplicity of presentation of the scheme hereafter. The relays use a unit-norm beamforming vector ${\bm d}\in\mathbb{C}^K$ during the second phase, i.e., the $k$th relay multiplies the transmitted signal $x_r$ by $d_k$ and sends it through the channel, and hence the received signals at the legitimate users and the eavesdropper are given by
\begin{align}
y^r_j&={\bm g_j}^\dagger{\bm d}x_r+n^r_j,\quad j=1,2,e
\end{align}
%\begin{align}
%y^r_1&={\bm g_1}^\dagger{\bm d}x_r+n^r_1\\
%y^r_2&={\bm g_2}^\dagger{\bm d}x_r+n^r_2\\
%y^r_e&={\bm g_e}^\dagger{\bm d}x_r+n^r_e
%\end{align}
where the superscript $r$ is to denote signals received from the relays, and the noise terms $n^r_1$, $n^r_2$, and $n^r_e$ are i.i.d. $\sim\mathcal{CN}\left(0,1\right)$. We let the relays use independent codewords from those used by the BS to forward their messages. The achieved rates at the legitimate users after the second phase are given by (\ref{eq_dec_rate_1}) and (\ref{eq_dec_rate_2}) at the top of the next page \cite[Theorem 16.2]{elgamalKim}, given that the relays follow the decoding order $(i)$, with the superscript $DF$ denoting decode-and-forward.
\begin{figure*}[t]
\begin{align}
r^{DF}_1&=\min\left\{\log\left(1+|h_1|^2\alpha\bar{P}\right)+\log\left(1+\left|{\bm g}_1^\dagger{\bm d}\right|^2\alpha\left(P-\bar{P}\right)\right),\min_{1\leq k\leq K}R_{k,1}^{(i)}\right\} \label{eq_dec_rate_1} \\
r^{DF}_2&=\min\left\{\log\left(1+\frac{|h_2|^2\bar{\alpha}\bar{P}}{1+|h_2|^2\alpha \bar{P}}\right)+\log\left(1+\frac{\left|{\bm g}_2^\dagger{\bm d}\right|^2\bar{\alpha}\left(P-\bar{P}\right)}{1+\left|{\bm g}_2^\dagger{\bm d}\right|^2\alpha \left(P-\bar{P}\right)}\right),\min_{1\leq k\leq K}R_{k,2}^{(i)}\right\} \label{eq_dec_rate_2}
\end{align}
\hrulefill
\vspace{-.2in}
\end{figure*}

For $K\geq2$, we design the beamforming vector ${\bm d}$ to be a unit-norm vector orthogonal to ${\bm g}_e$, and denote it by ${\bm d}_o$. This way, the eavesdropper does not gain any useful information during the second phase. Thus, we have
\begin{align} \label{eq_dec_null}
{\bm g}_e^\dagger{\bm d}_o=0.
\end{align}
Further, for $K\geq3$, we have that $\{{\bm g}_1,~{\bm g}_2,~{\bm g}_e\}$ are linearly independent w.p. $1$, and hence
\begin{align}
|{\bm g}_1^\dagger{\bm d}_o|>0,~|{\bm g}_2^\dagger{\bm d}_o|>0,\quad\text{w.p. $1$}.
\end{align}
Thus, the achievable secrecy rates in this case are given by
\begin{align}
r_{s,1}^{DF}=&\frac{1}{2}\left[r^{DF}_1-\log\left(1+|h_e|^2\alpha\bar{P}\right)\right]^+ \label{eq_sec_rate_dec_1} \\
r_{s,2}^{DF}=&\frac{1}{2}\left[r^{DF}_2-\log\left(1+\frac{|h_e|^2\bar{\alpha}\bar{P}}{1+|h_e|^2\alpha \bar{P}}\right)\right]^+ \label{eq_sec_rate_dec_2}
\end{align}
where the extra multiplication by $\frac{1}{2}$ is due to transmission of the same message over two phases with equal durations.

We now optimize the beamforming vector ${\bm d}_o$. Towards that end, we rewrite the constraint in (\ref{eq_dec_null}) slightly differently as
\begin{align} \label{eq_dec_null_alt}
{\bm d}_o=\mathcal{P}^\perp\!\left({\bm g}_e\right){\bm u}_d
\end{align}
where $\mathcal{P}^\perp\!\left(\cdot\right)$, as defined in (\ref{eq_p_perp_original}), now represents a projection matrix onto the orthogonal complement of vectors in $\mathbb{C}^K$, and ${\bm u}_d\in\mathbb{C}^K$ is some vector to be designed. Next, it is direct to see that $r^{DF}_1$ is non-decreasing in $\left|{\bm g}_1^\dagger{\bm d}_o\right|^2$ and that $r^{DF}_2$, after simple first derivative analysis, is non-decreasing in $\left|{\bm g}_2^\dagger{\bm d}_o\right|^2$. Hence, one needs to choose ${\bm d}_o$ to maximize these terms. We propose maximizing their convex combination $\beta\left|{\bm g}_1^\dagger{\bm d}_o\right|^2+(1-\beta)\left|{\bm g}_2^\dagger{\bm d}_o\right|^2$, for some $0\leq\beta\leq1$ of choice. Using (\ref{eq_dec_null_alt}), and after simple manipulations, the optimal $\hat{{\bm u}}_d$ is given by the solution of the following problem:
\begin{align} \label{opt_dec_beta}
\max_{{\bm u}_d}\quad&{\bm u}_d^\dagger\mathcal{P}^\perp\!\left({\bm g}_e\right)\left(\beta{\bm g}_1{\bm g}_1^\dagger+(1-\beta){\bm g}_2{\bm g}_2^\dagger\right)\mathcal{P}^\perp\!\left({\bm g}_e\right){\bm u}_d \nonumber \\
\mbox{s.t.}\quad&{\bm u}_d^\dagger\mathcal{P}^\perp\!\left({\bm g}_e\right){\bm u}_d=1
\end{align}
and therefore $\hat{{\bm u}}_d$ is given by the leading eigenvector of the (Hermitian) matrix:
\begin{align}
\mathcal{P}^\perp\!\left({\bm g}_e\right)\left(\beta{\bm g}_1{\bm g}_1^\dagger+(1-\beta){\bm g}_2{\bm g}_2^\dagger\right)\mathcal{P}^\perp\!\left({\bm g}_e\right)
\end{align}
i.e., the eigenvector corresponding to its largest eigenvalue. Finally, the optimal $\hat{{\bm d}}_o$ is given by
\begin{align}
\hat{{\bm d}}_o=\frac{\mathcal{P}^\perp\!\left({\bm g}_e\right)\hat{{\bm u}}_d}{\|\mathcal{P}^\perp\!\left({\bm g}_e\right)\hat{{\bm u}}_d\|}.
\end{align}
Substituting $\hat{{\bm d}}_o$ in (\ref{eq_sec_rate_dec_1}) and (\ref{eq_sec_rate_dec_2}) gives the secrecy rates.

%================================
\section{Amplify-and-Forward}

In this section, we discuss the amplify-and-forward scheme. As in the decode-and-forward scheme, communication takes place over two phases. In the first phase, the received signals at the legitimate users, eavesdropper, and relays are given by (\ref{eq_rec_sig_1})--(\ref{eq_rec_sig_2}), (\ref{eq_rec_sig_e}), and (\ref{eq_rec_sig_r}), respectively, with a total transmit power $\bar{P}\leq P$. In the second phase, the $k$th relay amplifies its received signal by multiplying it by a constant $a_k$ and sends it through the channel. Effectively, this can be written as the multiplication $\texttt{diag}\left({\bm a}\right){\bm y}_r$, where ${\bm a}\in\mathbb{C}^K$ is a beamforming vector to be designed, and $\texttt{diag}\left({\bm a}\right)$ is a diagonalization of the vector ${\bm a}$. The received signals at the legitimate users and the eavesdropper in the second phase are given by
\begin{align}
y^r_j={\bm g}_j^\dagger\texttt{diag}\left({\bm a}\right){\bm y}_r+n^r_j,\quad j=1,2,e.
\end{align}
%\begin{align}
%y^r_1={\bm g}_1^\dagger\texttt{diag}\left({\bm a}\right){\bm y}_r+n^r_1 \\
%y^r_2={\bm g}_2^\dagger\texttt{diag}\left({\bm a}\right){\bm y}_r+n^r_2 \\
%y^r_e={\bm g}_e^\dagger\texttt{diag}\left({\bm a}\right){\bm y}_r+n^r_e
%\end{align}
Now observe that from, e.g., the strong user's perspective, this amplify-and-forward scheme can be viewed, using (\ref{eq_rec_sig_r}), as the following single-input multiple-output (SIMO) system:
\begin{align}
\begin{bmatrix}y_1\\y^r_1\end{bmatrix}=\begin{bmatrix}h_1\\{\bm g}_1^\dagger\texttt{diag}\left({\bm a}\right){\bm h}_r\end{bmatrix}x+\begin{bmatrix}n_1\\\tilde{n}^r_1\end{bmatrix}
\end{align}
where the noise term $\tilde{n}^r_1\triangleq{\bm g}_1^\dagger\texttt{diag}\left({\bm a}\right){\bm n}_r+n^r_1$ is complex-Gaussian with zero mean and variance $\mathbb{E}\left[|\tilde{n}^r_1|^2\right]={\bm g}_1^\dagger\texttt{diag}\left({\bm a}^*\right)\texttt{diag}\left({\bm a}\right){\bm g}_1+1$, with the superscript $*$ denoting the conjugate operation. One can write similar equations for the weak user as well. Hence, the achievable rates at the legitimate users of this SIMO system after the second phase are given by \cite[Section 5.3.1]{tse-wireless}
\begin{align}
r^{AF}_1&\!\!=\!\log\!\left(\!1\!+|h_1|^2\alpha\bar{P}+\frac{{\bm a}^\dagger{\bm G}_{1,r}{\bm a}}{1+{\bm a}^\dagger{\bm G}_1{\bm a}}\alpha\bar{P}\!\right) \label{eq_rate_amp_1} \\
r^{AF}_2&\!\!=\!\log\!\left(\!1\!+\frac{|h_2|^2\bar{\alpha}\bar{P}}{1\!+\!|h_2|^2\alpha\bar{P}}+\frac{{\bm a}^\dagger{\bm G}_{2,r}{\bm a}\bar{\alpha}\bar{P}}{1\!+\!{\bm a}^\dagger{\bm G}_2{\bm a}\!+\!{\bm a}^\dagger{\bm G}_{2,r}{\bm a}\alpha\bar{P}}\!\right) \label{eq_rate_amp_2}
\end{align}
with the superscript $AF$ denoting amplify-and-forward, and
\begin{align}
{\bm G}_{j,r}&\triangleq\texttt{diag}\left({\bm h}_r^*\right){\bm g}_j{\bm g}_j^\dagger\texttt{diag}\left({\bm h}_r\right),\quad j=1,2 \\
{\bm G}_j&\triangleq\texttt{diag}\left({\bm g}_j^*\right)\texttt{diag}\left({\bm g}_j\right),\quad j=1,2.
\end{align}

As for the eavesdropper, observe that by (\ref{eq_rec_sig_r}) we have
\begin{align}
{\bm g}_e^\dagger\texttt{diag}\left({\bm a}\right){\bm y}_r={\bm g}_e^\dagger\texttt{diag}\left({\bm a}\right){\bm h}_rx+{\bm g}_e^\dagger\texttt{diag}\left({\bm a}\right){\bm n}_r.
\end{align}
Upon noting that ${\bm g}_e^\dagger\texttt{diag}\left({\bm a}\right){\bm h}_r={\bm g}_e^\dagger\texttt{diag}\left({\bm h}_r\right){\bm a}$, we propose, for $K\geq2$, designing the beamforming vector ${\bm a}$ to be orthogonal to the vector $\texttt{diag}\left({\bm h}_r^*\right){\bm g}_e$ and denote it by ${\bm a}_o$. This way, the eavesdropper does not gain any useful information during the second phase. Thus, we have
\begin{align} \label{eq_amp_null}
{\bm g}_e^\dagger\texttt{diag}\left({\bm h}_r\right){\bm a}_o=0.
\end{align}
As in the decode-and-forward scheme, we have for $K\geq3$ that $\{{\bm g}_1,~{\bm g}_2,~{\bm g}_e\}$ are linearly independent w.p. $1$. Hence
\begin{align}
|{\bm g}_1^\dagger\texttt{diag}\left({\bm h}_r\right){\bm a}_o|>0,~|{\bm g}_2^\dagger\texttt{diag}\left({\bm h}_r\right){\bm a}_o|>0,~\text{w.p. $1$}.
\end{align}
Thus, the achievable secrecy rates in this case are given by
\begin{align}
r_{s,1}^{AF}=&\frac{1}{2}\left[r^{AF}_1-\log\left(1+|h_e|^2\alpha\bar{P}\right)\right]^+ \label{eq_sec_rate_amp_1} \\
r_{s,2}^{AF}=&\frac{1}{2}\left[r^{AF}_2-\log\left(1+\frac{|h_e|^2\left(1-\alpha\right)\bar{P}}{1+|h_e|^2\alpha \bar{P}}\right)\right]^+ \label{eq_sec_rate_amp_2}
\end{align}
where the extra multiplication by $\frac{1}{2}$ is due to transmission of the same message over two phases of equal durations.

We now focus on further optimizing the beamforming signal ${\bm a}_o$. Towards that, we first note that the power transmitted in the second phase by the relays is given by
\begin{align}
&\mathbb{E}\left[{\bm a}_o^\dagger\texttt{diag}\left({\bm y}_r^*\right)\texttt{diag}\left({\bm y}_r\right){\bm a}_o\right] \nonumber \\
&\hspace{.05in}={\bm a}_o^\dagger\left(\texttt{diag}\left({\bm h}_r^*\right)\texttt{diag}\left({\bm h}_r\right)\bar{P}+{\bm I}_K\right){\bm a}_o\triangleq{\bm a}_o^\dagger{\bm A}{\bm a}_o. \label{eq_amp_pwr}
\end{align}
Next, we rewrite the constraint (\ref{eq_amp_null}) slightly differently as
\begin{align} \label{eq_amp_null_alt}
{\bm a}_o=\mathcal{P}^\perp\!\left(\texttt{diag}\left({\bm h}_r\right){\bm g}_e\right){\bm u}_a\triangleq{\bm F}{\bm u}_a
\end{align}
for some vector ${\bm u}_a\in\mathbb{C}^K$ to be designed. Next, we note that for the strong user, using (\ref{eq_amp_pwr}) and (\ref{eq_amp_null_alt}), finding the optimal ${\bm u}_a$ is tantamount to solving the following problem (note that ${\bm F}$ is a Hermitian matrix):
\begin{align}
\max_{{\bm u}_a}\quad&\frac{{\bm u}_a^\dagger{\bm F}{\bm G}_{1,r}{\bm F}{\bm u}_a}{1+{\bm u}_a^\dagger{\bm F}{\bm G}_1{\bm F}{\bm u}_a} \nonumber \\
\mbox{s.t.}\quad&{\bm u}_a^\dagger{\bm F}{\bm A}{\bm F}{\bm u}_a\!=\!P-\bar{P}
\end{align}
which can be equivalently rewritten as the following problem:
\begin{align}
\max_{{\bm u}_a}\quad&\frac{{\bm u}_a^\dagger{\bm F}{\bm G}_{1,r}{\bm F}{\bm u}_a}{{\bm u}_a^\dagger{\bm F}\left(\frac{1}{P-\bar{P}}{\bm A}+{\bm G}_1\right){\bm F}{\bm u}_a}
\end{align}
whose solution is given by the leading {\it generalized} eigenvector \cite{mtrx-comp} of the following matrix pencil:
\begin{align}
\left({\bm F}{\bm G}_{1,r}{\bm F}~,~{\bm F}\left(\frac{1}{P-\bar{P}}{\bm A}+{\bm G}_1\right){\bm F}\right)
\end{align}
i.e., the generalized eigenvector corresponding to the largest generalized eigenvalue of the pencil. Let us denote such a vector by ${\bm u}_a^{(1)}$. Similarly, one can show that the optimal ${\bm u}_a$ for the weak user is given by the leading generalized eigenvector of the following matrix pencil:
\begin{align}
\hspace{-.1in}\left(\!{\bm F}{\bm G}_{2,r}{\bm F}~,~{\bm F}\left(\frac{1}{P-\bar{P}}{\bm A}+{\bm G}_2+{\bm G}_{2,r}\alpha\bar{P}\right){\bm F}\!\right)
\end{align}
which we denote by ${\bm u}_a^{(2)}$. To satisfy the power constraint (\ref{eq_amp_pwr}) and the orthogonality constraint (\ref{eq_amp_null_alt}), the corresponding ${\bm a}_o^{(j)}$, $j=1,2$, is given by
\begin{align}
{\bm a}_o^{(j)}=\sqrt{\frac{P-\bar{P}}{{\bm u}_a^{(j)T}{\bm F}{\bm A}{\bm F}{\bm u}_a^{(j)}}}{\bm F}{\bm u}_a^{(j)},\quad j=1,2.
\end{align}
Then, as in the decode-and-forward scheme, we propose choosing the optimal $\hat{{\bm a}}_o$ by the following convex combination:
\begin{align} \label{opt_amp_beta}
\hat{{\bm a}}_o=\beta{\bm a}_o^{(1)}+(1-\beta){\bm a}_o^{(2)}
\end{align}
for some $0\leq\beta\leq1$ of choice. Given $\hat{{\bm a}}_o$, we substitute in (\ref{eq_sec_rate_amp_1}) and (\ref{eq_sec_rate_amp_2}) to get the achievable secrecy rates.

%================================
\section{Numerical Evaluations} \label{sec_num}

In this section, we first discuss evaluating the optimal transmit power and the optimal power fraction for each of the proposed relaying schemes, such that the secrecy rate region is maximized. Specifically, we characterize the boundary of the secrecy rate region by solving the following problem:
\begin{align}
\max_{\alpha,\bar{P}}\quad&\mu r_{s,1}^n+(1-\mu)r_{s,2}^n \nonumber \\
\mbox{s.t.}\quad&0\leq\bar{P}\leq P,\quad0\leq\alpha\leq1
\end{align}
for some $\mu\in[0,1]$, and $n\in\{J,DF,AF\}$ corresponds to one of the proposed schemes. For the decode-and-forward scheme, we do an outer maximization over the decoding order $(i)$, $i=1,2$ (see (\ref{eq_dec_rate_1}) and (\ref{eq_dec_rate_2})). We use a line search algorithm to numerically solve the above problem. Note that the feasible set is bounded, which facilitates the convergence of the algorithm to an optimal point. We use $\beta=\mu$ in (\ref{opt_dec_beta}) and (\ref{opt_amp_beta}).

The physical layout we consider is a one-dimensional system, where the strong user, the weak user, and the eavesdropper are located at $30$ meters, $40$ meters, and $50$ meters from the BS, respectively. We have $K=5$ relays, and for simplicity we assume that they are all close enough to each other that they are approximately at the same distance of $15$ meters from the BS. To emphasize the effect of distance on the channel gains, we use the following simplified channel model \cite{petropulu-coop-relay-security}: $h=\sqrt{1/l^\gamma}e^{j\theta}$, where $h$ is the channel gain between two nodes, $l$ is the distance between them, $\gamma=3.5$ is the path loss exponent, and $\theta$ is a uniform random variable in $[0,2\pi]$. We denote by $l_e$ the distance from the BS to the eavesdropper. We set $P$ to $30$ dBm\footnote{Note that the noise power is normalized in this paper, and hence $P$ also represents the SNR.}. We run multiple iterations of the simulations and compute the average performance.

\begin{figure}[t]
\center
\includegraphics[scale=.45]{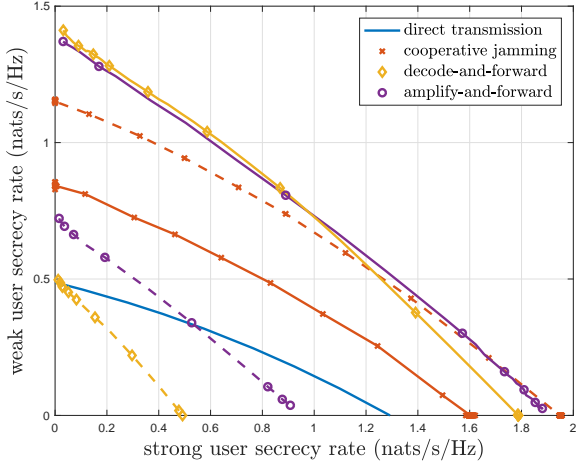}
\caption{Achievable secrecy rate regions of the proposed schemes. Solid lines are when $l_e=50$ meters, and dashed lines are when $l_e=20$ meters.}
\label{fig_sec_reg_all_schemes}
\vspace{-.1in}
\end{figure}

In Fig.~\ref{fig_sec_reg_all_schemes}, we plot the achievable secrecy rate regions of the proposed schemes. Solid lines represent the system parameters stated above, and dashed lines represent the case of $l_e=20$ meters. We see that when $l_e=50$ meters, cooperative jamming outperforms direct transmission and is in turn outperformed by both decode-and-forward and amplify-and-forward which perform relatively close to each other. With $l_e=20$ meters, direct transmission achieves zero secrecy rates since the eavesdropper is closer to the BS than both users. However, strictly positive secrecy rates are achievable by all the relaying schemes. We also see that cooperative jamming performs best, since the relays are close to the eavesdropper and hence their jamming effect is quite powerful. For parts of the region, it even performs very close to decode-and-forward and amplify-and-forward for the $l_e=50$ meters case.

\begin{figure}[t]
\center
\includegraphics[scale=.45]{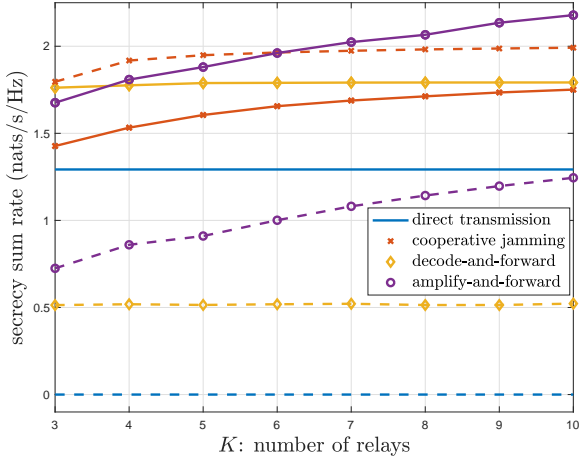}
\caption{Achievable secrecy sum rates vs. the number of relays. Solid lines are when $l_e=50$ meters, and dashed lines are when $l_e=20$ meters.}
\label{fig_sum_sec_rate_nmbr_relays}
\vspace{-.1in}
\end{figure}

Next, we show the effect of the number of relays on the achievable {\it secrecy sum rates} of the proposed schemes in Fig.~\ref{fig_sum_sec_rate_nmbr_relays}. Again we observe that all relaying schemes achieve positive secrecy sum rates when $l_e=20$ meters and that cooperative jamming performs best in this case. We also observe that amplify-and-forward is more sensitive than the other schemes to the number of relays.

\begin{figure}[t]
\center
\includegraphics[scale=.45]{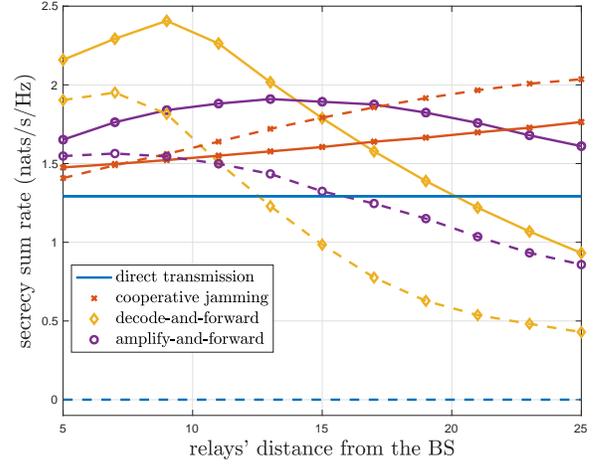}
\caption{Achievable secrecy sum rates vs. relays' distance from the BS. Solid lines are when $l_e=50$ meters, and dashed lines are when $l_e=27$ meters.}
\label{fig_sum_sec_rate_relays_distance}
\vspace{-.1in}
\end{figure}

Finally, we fix the users' distances and show the effect of the relays' distance from the BS (and hence the users) on the achievable secrecy sum rates in Fig.~\ref{fig_sum_sec_rate_relays_distance}. For this case, we vary the relays' distance but still keep them closer to the BS than the legitimate users and the eavesdropper. The dashed lines in this case are when $l_e=27$ meters. We see that only the cooperative jamming scheme's performance monotonically increases with the relays' distance, which is again attributed to the fact that the jamming effect is more powerful when the relays get closer to the eavesdropper.

%================================
\section{Conclusion}

The security benefits of using trusted cooperative half duplex relays in a two-user SISO NOMA downlink system with an external eavesdropper have been studied under three relaying schemes: cooperative jamming, decode-and-forward, and amplify-and-forward. For each scheme, secure beamforming signals have been designed at the relays to maximize the achievable secrecy rate region. While the schemes improve the secrecy rate region, their performance varies depending on the eavesdropper's and relays' distances from the BS and also on the number of relays.

%================================
\bibliographystyle{unsrt}

\end{document}